# Directly Visualizing the Crossover from Incoherent to Coherent Phonons in Two-dimensional Periodic MoS$_2$/MoSe$_2$ Arrayed Heterostructure

Meng An[1, 2], Dongsheng Chen[1], Weigang Ma[2,*], Shiqian Hu[3,*], Xing Zhang[2]

[1]College of Mechanical & Electrical Engineering, Shaanxi University of Science and Technology, Xi'an 710021, People's Republic of China

[2]Key Laboratory for Thermal Science and Power Engineering of Ministry of Education, Department of Engineering Mechanics, Tsinghua University, Beijing 100084, China

[3]Department of Physics, Yunnan University, Kunming 650091, China

[*]Corresponding email:

maweigang@tsinghua.edu.cn (W. G. M.)

shiqian@ynu.edu.cn  (S. Q. H.)

**Abstract**

Recently, massive efforts have been done on controlling thermal transport via coherent phonons in the various periodic nanostructures. However, the intrinsic lattice difference between the constituent materials inevitably generates the disorder at the interfaces, thus limiting the opportunity of directly observing the coherent phonon transport. Here, we investigate the controllability and visualization of the coherent phonon transport in a periodic $MoS_2/MoSe_2$ arrayed heterostructure with minimum lattice mismatching using non-equilibrium molecular dynamics simulation. It is found that the coherent phonon transport can be destroyed and rebuilt through adjusting the density of $MoSe_2$ nanodot arrays. The phonon localization induced by the destruction of correlation is visualized based on the spatial energy distribution and anharmonic analysis. Furthermore, the *eigen* vector diagrams provide a distinct visualization of the localized phonon modes. Besides, the correlation of phonon can be rebuilt by reducing the period length, which is verified by the enhanced group velocities extracted from phonon dispersion curves. Interestingly, the crossover from incoherent to coherent phonon transport is directly observed by the spatial energy distributions and the spectral phonon transmission coefficients. Finally, the size and temperature dependence of thermal conductivity are also discussed. This study of the phonon coherence and its visualizing manipulation on thermal conductivity will be beneficial to fine heat control and management in the real applications.

**Introduction**

Two-dimensional (2D) materials have drawn growing research interests due to their promising potential in electronics[1], optoelectronics[2] and thermoelectrics[3]. Furthermore, both $MoS_2$[4] and $MoSe_2$[5] show the transition from indirect-gap to direct-gap in the visible range of wavelengths when the dimension of the sample reduces from bulk to monolayer. Alternatively, a direct band gap can also be realized in the 2D heterostructures from artificial assembly of various layered 2D materials with alternating stacking orders, such as bilayer $MoSe_2/WSe_2$ lateral heterostructures[6]. Over the past years, rapid development of synthesis methods, such as mechanical exfoliation[7], chemical vapor deposition (CVD)[8] and molecular beam epitaxy (MBE)[9] promote their extraordinary applications. Particularly, the synthesized periodic $MoS_2/MoSe_2$ arrayed heterostructures (PAHs)[10-12] dramatically enhanced the integrity of microstructures by controlling CVD growth and transferring arrayed $MoS_2$ monolayers on predetermined locations in a unique process[11]. Such promising integrated electronics systems not only can avoid the contaminations and structure damage during the transfer process, but be help to fulfill their excellent performance of electronic devices. In addition, the engineering phonon transport properties of 2D materials such as patterned nanostructures and van der Waals stacked forms can also be utilized to improve thermoelectric performance[13, 14]. However, the sample preparation and thermal properties measurement of nanostructured 2D materials are still extremely challenging[14, 15], but under continuous development. Studies to date are largely limited to electron-related fundamental science and diverse device technology while few works have paid attention to its thermal properties of PAHs that is beneficial for large-scale practical applications. Therefore, to explore more potential of PAHs-based electronic devices at a suitable temperature, it is highly desirable to understand the thermal transport properties of PAHs. In particular, the triangular structure is more preferred by experiment researchers[10, 15], thus, we mainly focus on the study of the triangular PAHs in this work.

Phonons, quanta of lattice vibrations, have wave-particle duality, which governs the thermal transport of semiconductor and dielectric materials, *i.e.*, the particle-like incoherent phonons that appear when the phonons lose their phase after successive phonon scattering events, and the wave-like coherent phonons that appear when the phase is preserved. Prior studies on the thermal transport of incoherent and coherent phonon mainly focused on the demonstration[16, 17] and manipulation[18-20] of incoherent and coherent phonons in various superlattices and metamaterials. For example, Gang Chen et al.[16] observed the coherent phonons heat conduction in superlattices by experimental measurements. Ravichandran et al.[17] presented an unambiguous demonstration of crossover from particle-like (incoherent) to wave-like (coherent) phonon in high-quality epitaxial perovskite oxide superlattices. Besides, the influence of disorder on the coherent phonon transport was studied with different porosities at room temperature in a 2D $C_3N$ phononic crystal structure[18]. The phonon transmission revealed that the low-frequency coherent phonons were largely suppressed due to the introduction of disorder. Likewise, the coherent phonons can also be controlled by the period of the 2D lateral superlattice[19, 20]. For instance, Wang et al.[19] found that the thermal conductivities of silicene-germanene superlattice first reduces and then increases with the increasing period length, where a minimum value is obtained, which suggests the phonon transport transition from coherent to incoherent mechanism. However, the understanding of phonon transport from incoherent to coherent mechanism is still elusive for PAHs widely used in various electron-related devices.

Besides, direct visualization of incoherent and coherent phonon propagation can promote deeper understanding and design thermal management of PAHs. With the development of temporal and spatial resolution of the experimental techniques, several groups have detected coherent phonons under different additional conditions[21, 22]. However, the direct visualizing and analysis at mode-level of the crossover from incoherent to coherent phonon transport is still unclear. To address this issue, a computational technique such as molecular dynamics (MD) simulations can complement experimental effort and provide critical insights into phonon transport at the atomic level. In our study, we investigated the thermal conductivity of PAHs using no-equilibrium molecular dynamics (NEMD) simulations. The thermal conductivities of PAHs with the different $MoSe_2$ nanodots densities were studied in details. The mode weight factor and *eigen* vector diagrams were used to visualize the localization mechanism of incoherent phonons. Moreover, the transition from incoherent to coherent phonons transport was identified by enhanced phonon group velocities, which was further explained by the visual-spatial energy distributions and phonon spectral transmission coefficients. Besides, the size and temperature dependence of $MoS_2/MoSe_2$ PAHs were discussed.

## NEMD Method

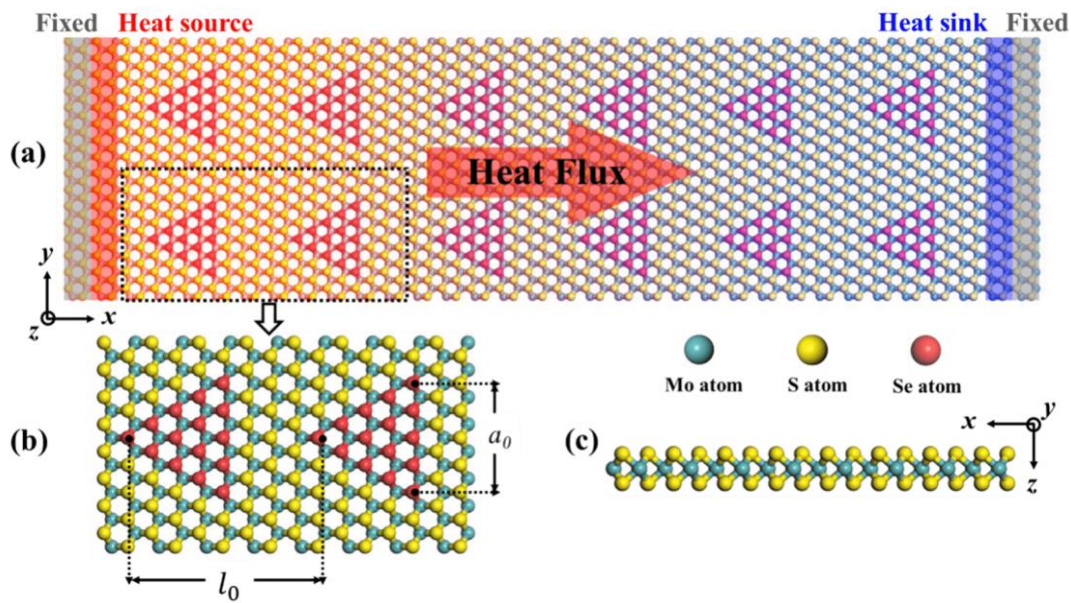

**Figure 1.** (a) Schematic picture of NEMD simulations for calculating the thermal conductivity of single-layer $MoS_2/MoSe_2$ PAHs sheet. The enlarged picture of two primitive cells in the black dotted box is shown in (b) and that of the side view is shown in (c), where the triangular $MoSe_2$ nanodot is embedded in the $MoS_2$ sheet. The $l_0$ and $a_0$ represent the length of the primitive cell in the $x$ direction and the side length of triangular $MoSe_2$ nanodots, respectively.

In the current work, all of the NEMD simulations were implemented by large-scale atomic/molecular massively parallel simulator (LAMMPS) package[23]. The Stillinger-Weber (SW) is adopted to calculate the thermal conductivity of PAHs. To show the reliability of MD results, REBO empirical potential is also compared to check the thermal conductivity calculation of pristine $MoS_2$ (details in supplementary information **Table S2**). The simulation results show that the SW potential can reproduce the thermal conductivity of experimental measurement[24] and the corresponding computational cost is relatively cheaper compared with REBO potential. In addition, the SW potential have also been utilized to study the influence of defect concentration on thermal conductivity of $MoS_2$. Moreover, the emphasis of this work is to explore the thermal transport mechanism of 2D $MoS_2/MoSe_2$ nanodots PAHs. Thus, the SW potential [25, 26] is employed to describe the interatomic interactions in $MoS_2/MoSe_2$ nanodots PAHs. The parameters of SW force field developed by Jiang[26] were adopted in all the simulations (details in supplementary information **Table S1**), which have been successfully reproduced the thermal[27] and mechanical[26, 27] properties of transition metal dichalcogenides (TMDs) heterostructures. The schematic picture of $MoS_2/MoSe_2$ PAHs is shown in **Figure 1**. The system sizes are fixed at 54.02 nm $\times$ 4.99 nm. The thickness of the system is set as 0.61 nm. $l_0$ is the period length of the unit cell in $x$ direction. The side length $a_0$ of the triangular nanodots is fixed at 1.25 nm shown in **Figure 1b**. The thermal conductivity of different period lengths $l_0$ is calculated for the fixed the system length 54.02 nm. To establish a steady temperature gradient along $x$ direction, the system was coupled with Langevin thermostats[28] at the 3rd to 5th and (N-5)th to (N-3)th layers with 310 and 290 K, respectively in in Figure 1a. Atoms at the boundaries (from the 1th to 2nd and (N-1)th to Nth layers) are fixed to simulate adiabatic walls. The periodic boundary conditions were applied in the zigzag ($y$) direction.

The thermal conductivity $\kappa$ of the single-layer $MoS_2/MoSe_2$ PAHs was calculated based on the Fourier's Law[29],

$$\kappa = -\frac{Q}{A \cdot \nabla T} \qquad (1)$$

where $Q$ and $\nabla T$ are the heat flux and the temperature gradient ($\nabla T = \Delta T / \Delta L$, where $\Delta L$ and $\Delta T$ are the length of system and temperature difference in $x$ direction, respectively, where the temperatures of fixed and heat bath regions are excluded.). *A* is the cross-sectional area. The velocity Verlet algorithm[30] is adopted to integrate the discrete differential equation of motions with a time step 0.5 fs. At first, the energy minimization was performed to relax the residual stress between the atoms. Then, the initial relaxed configuration was equilibrated in the isothermal-isobaric (NPT) and canonical (NVT) ensemble for 0.5 ns, respectively. After the NVT relaxation, the microcanonical (NVE) ensemble was performed for 10.5 ns to reach the steady-state, where the energy and temperature profile are recorded to calculate thermal conductivity for the last 10 ns. The results presented here are averaged over five independent simulations with different initial velocities conditions.

Intrinsically, molecular dynamics (MD) simulations alone do not describe the wave-like phonon transport, which can present atomic displacement field based on the empirical interatomic potential and masses of a structure in dynamical equilibrium or nonequilibrium. Combined with lattice dynamics (LD), MD simulations can sample the vibrational phase space of atoms in terms of the vibrational modes (*i.e.* phonons in the case of perfect crystal) and give insights to the vibrational mode level contributions to thermal transport, which can include wave-like properties of phonon, such as coherent phonon and resonance[31, 32]. In this work, MD simulation combined with lattice dynamics was utilized to explore the wave-like behaviors of phonons. The vibration vector of phonons can be calculated by the lattice dynamic package "general utility lattice program" (GULP)[33].

**Results and Discussions**

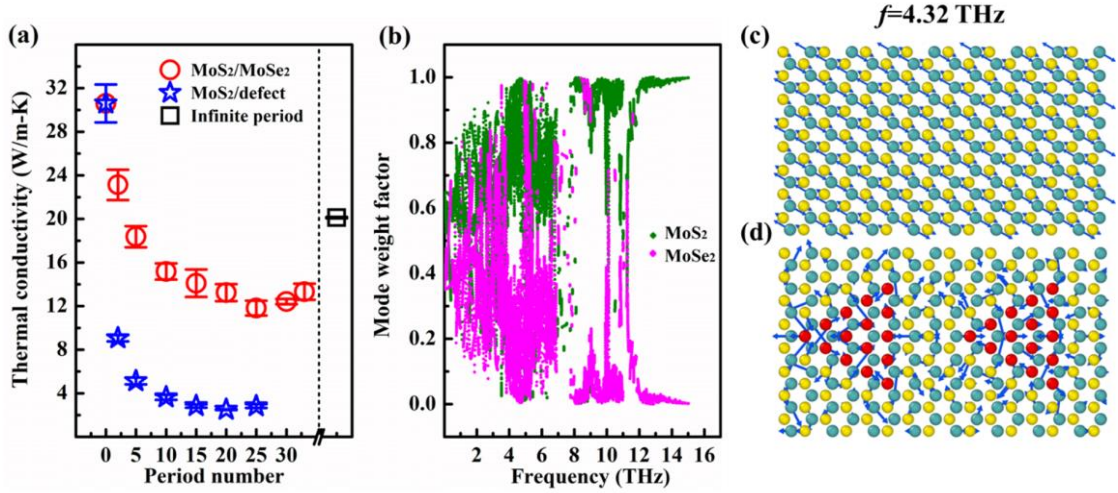

**Figure 2.** (Color online) (a) The thermal conductivities of $MoS_2/MoSe_2$ (red circles) and $MoS_2$/defect (blue stars) PAHs as a function of the period number along heat flux transported direction. (b) Phonon mode weight factor of $MoS_2$ and $MoSe_2$ region in PAHs with period number 25 ($l_0$=2.16 nm). (c-d) The eigen vector diagrams of pristine $MoS_2$ and $MoS_2/MoSe_2$ PAHs at $f$ = 4.32 THz.

This study mainly focuses on the lattice thermal conductivity of single-layer $MoS_2/MoSe_2$ PAHs. The thermal conductivity as a function of the period number of embedded $MoSe_2$ nanodots is shown in **Figure 2a**, where the system length along heat energy transported and the side length of triangle $MoSe_2$ nanodots are fixed as 54.02 nm and 1.25 nm, respectively. The calculated thermal conductivities of pristine single-layer $MoS_2$ are 30.60 and 12.90 W/m-K for the system size L=54.02 and 20.53 nm at room temperature, which is comparable with the previous results at the same size (details in supplementary information **Table S3**). For example, Ali Rajabpour et al. [34] obtained the thermal conductivity of monolayer MoS2 with system length L=20.22 nm is around 4.00 W/m-K. Moreover, it is found that the width is insensitive to the thermal conductivity of $MoS_2/MoSe_2$ PAHs (details in supplementary information **Figure S7**). In addition, previous studies have demonstrated that the asymmetric nanoscale systems such as triangle graphene nanoribbon, can realize the phenomenon of thermal rectification[35]. To check the thermal rectification of PAHs, the thermal conductivities of PAHs from left (right) to right (left) side are calculated, it is found the ratio of thermal rectification is around 1.03 for different periodic numbers, which can be negligible compared with other MD results of asymmetric systems[36, 37]. The details are shown in supplementary information **Table S4**.

For a fixed system length of 54.02 nm, the maximum period number of $MoSe_2$ nanodots is 33 because the fixed side length of triangle nanodots is 1.25 nm. It is observed that the thermal conductivity is reduced with an increase in the period number of embedded $MoSe_2$ nanodots (red circles), where the thermal conductivity reaches the lowest value of 11.82 W/m-K at the period number of 25, indicating that the density of embedded nanodots can serve as a robust way to enhance phonon scattering and block heat conduction. Besides, the $MoS_2$/defect PAHs is obtained by removing the $MoSe_2$-nanodots, and its thermal conductivity are suppressed to 2.83 W/m-K at the period number of 25, which indicates the defect-nanodots can efficiently block the phonon transport due to the enhanced coupling between low-frequency and high-frequency phonons and defect scatterings induced by the introduction of defect nanodots in our previous study[38]. Interestingly, the thermal conductivity of $MoS_2$/$MoSe_2$ PAHs shows an upward trend when the period number is larger than 25 (*i.e.* the period number ranging from 25 to 33). Furthermore, we build a PAHs with infinite period number, *i.e.*, the $l_0$ approaches zero and found the corresponding thermal conductivity (black rectangle) reaches up to 20.12 W/m-K, which further confirms the upward trend of thermal conductivity. This phenomenon can be explained: most of phonons that contributed to the thermal conductivity in PAHs ballistically traversed the system and hence were coherent. Most recently, a strongly coupled coherent phonons in single-layer $MoS_2$ was generated and detected by using a transient absorption setup with high temporal resolution[39]. However, as the period number increases, some phonons will be scattered by the increased interfaces of the embedded $MoSe_2$ nanodots. Previous studies demonstrated that interface roughness was effective in destroying the coherence of high-frequency phonons[16, 40]. The large reduction in thermal conductivity resulted from the loss of coherence of high-frequency phonons, but the low-frequency phonons that contribute to the thermal conductivity were mostly coherent during their transport through the single-layer structure until they are scattered at the sample boundaries. Above the critical period, the large period length provides enough space for the occurrence of phonon-phonon scattering before a phonon traverses the embedded nanodots in each unit, which can prevent the formation of coherent phonon modes. The observation of coherent effect necessitates the study of a short-period single-layer system. Such the transition from incoherent phonon transport to coherent phonon transport can also be found in many superlattices[17, 19], and phononic crystals[18].

In general, the reduction of thermal conductivity mainly originates from harmonic (phonon dispersion) and anharmonic effects arising from enhanced scattering between phonon modes, such as isotopes scattering[41], impurity scattering[42], and boundary scattering[43], *etc*. In our study, the embedded $MoSe_2$ nanodots create a static disorder that disrupts the lattice periodicity and the nature of phonon vibrational modes, which potentially enhances the coupling among phonon modes, increases phonon scattering and decreases phonon modes lifetime. To understand the reduction of thermal conductivity, we applied lattice dynamics to study the harmonic effect of the representative sample of the period number 25, where the minimum thermal conductivity is achieved. The mode weight factor (MWF) can be used to characterize the $MoS_2$ and $MoSe_2$ regional contribution of modes in $MoS_2/MoSe_2$ PAHs, and it is defined as[44, 45]

$$f_{j',\omega} = \sum_{j'} \sum_{\alpha} (\varepsilon_{j\alpha,\omega})^2 \qquad (2)$$

where the prime denotes that sum over $j$ (atoms) and is alternatively restricted to atoms of $MoS_2$ and $MoSe_2$. Therefore, the sum of the mode weight factors of $MoS_2$ and $MoSe_2$ regions in $MoS_2/MoSe_2$ nanodots PAHs is equal to unity, *i.e.* $f_{MoS_2,\omega} + f_{MoSe_2,\omega} = 1$. For a small MWF value, it represents a localized phonon mode, while a large MWF value corresponds to extended phonon modes, which contributes to thermal conduction. **Figure 2b** shows the MWF of both $MoS_2$ and $MoSe_2$ regions in one unit cell for all phonon modes. These phonon modes with frequency within the range [0,15 THz] are plotted. In thermal transport, these low-frequency phonons contribution [0-7 THz] are larger than that of high-frequency phonons [8-15 THz] due to their larger group velocity. It is found that these phonon modes MWF of the $MoSe_2$ region are far smaller than that of the $MoS_2$ region in the low-frequency phonons ranging from 4 to 7

THz, indicating that there are more localized modes in MoSe$_2$ region, while more extended modes in MoS$_2$ region. To observe the localized modes, we plotted the visualization of *eigen* vectors at the low frequency *f*=4.32 THz where the largest MWF difference is observed. Two typical examples (pristine MoS$_2$ and PAHs with period number 25) of *eigen* vectors at *f*=4.32 THz are plotted in **Figure 2c-d**. The blue arrows attached on each atom represent the atomic vibrational component in the phonon mode. The pristine MoS$_2$ sample in **Figure 2c** illustrates an ordered vibrational component, which indicate a delocalized mode. In contrast, atoms within the MoSe$_2$ region have large disorder vibrational component, while other atoms far away from the MoSe$_2$ regions have less vibrational component in **Figure 2d**, which indicates a highly localized mode in PAHs. From the inspection of *eigen* modes, it is observed that there is no phonon localization when atoms of oscillating in phase in **Figure 2c**. On the other hand, when the atomic vibrational directions are out of phase, the vibration mismatch between MoS$_2$ and MoSe$_2$ region leads to phonon localization, revealed by the spatial energy distribution of **Figure 4b** that phonons are localized in the triangular MoSe$_2$ nanodots due to phonon-phonon scattering and phonon-boundary scattering.

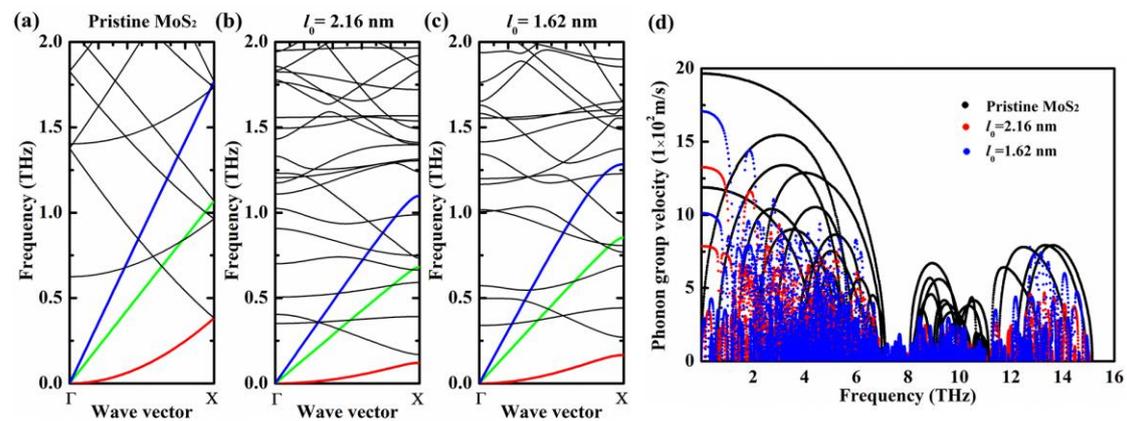

**Figure 3.** (a-c) Phonon dispersion relations of pristine MoS$_2$ and PAHs with period number 25 ($l_0$=2.16 nm) and period number 33 ($l_0$=1.62 nm), respectively. (d) The corresponding phonon group velocities of three systems in Figure 3(a-c).

To understand the harmonic effect of thermal conductivity with the increasing period number, the phonon dispersion curves of pristine MoS$_2$ and MoS$_2$/MoSe$_2$ PAHs with period number of 25 and 33 are shown in **Figure 3a-c**. On one hand, comparing to the pristine $MoS_2$, the phonon branches of TA, LA and ZA are significantly suppressed in PAHs of period number 25. Moreover, band gaps in the phonon dispersion are opened at the Brillouin boundaries, which is one of the typical phenomena in nanostructured phononic systems[45], indicating that wave interference and Bragg scattering are presented. Therefore, there will be many flatted phonon bands generated. On the other hand, with the increasing period number, *i.e.* 33 ($l_0$=1.62 nm), acoustic phonon branches become more stiffened, suggesting more phonons can transverse the PAHs system. Previous studies[16, 20] demonstrated the wave interference of phonon induces the existence of coherent phonon transport. To further quantitatively characterize the phonon transport, these group velocities of three typical systems are calculated as shown in **Figure 3d**. Their average group velocities are 2.35 km/s pristine $MoS_2$ (black dots), 1.36 km/s of PAHs with $l_0$=2.16 nm (red dots), and 1.40 km/s of PAHs with $l_0$=1.62 nm (blue dots) at 0-1 THz, respectively. This trend of group velocity with period number is consistent with the change of thermal conductivity, which further demonstrates that the transition from incoherent phonons to coherent phonons thermal transport in $MoS_2/MoSe_2$ PAHs.

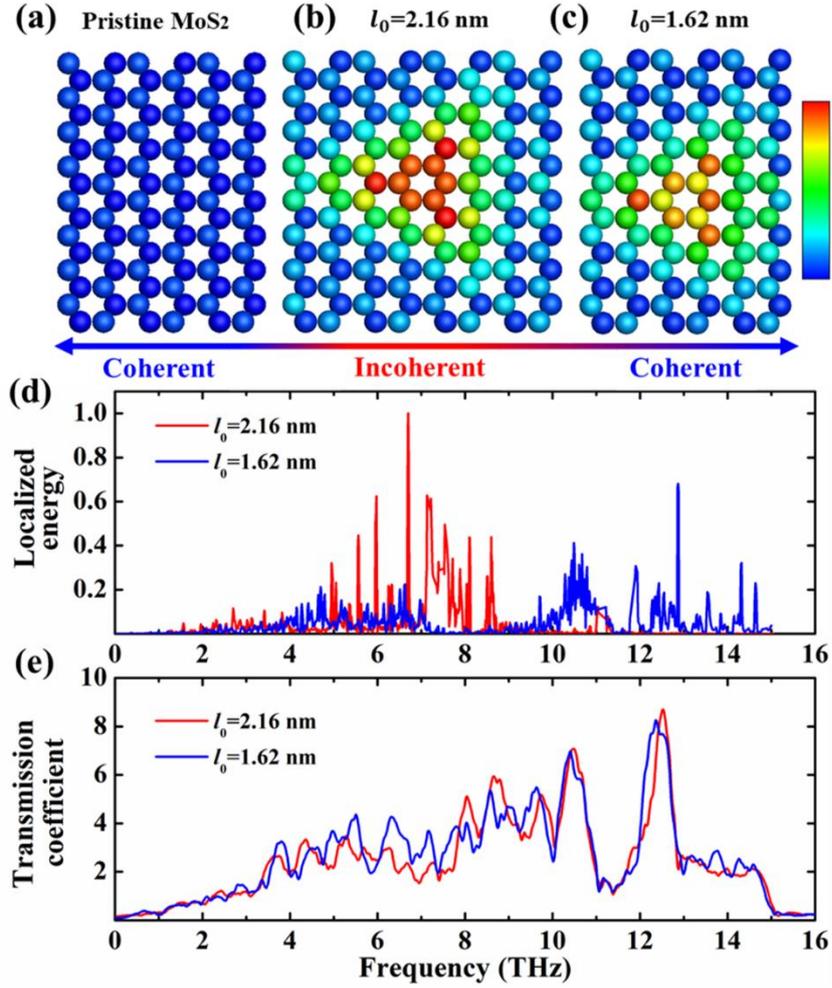

**Figure 4.** (a-c) Normalized spatial energy distributions of pristine MoS₂, $l_0$=2.16 nm (red line) and 1.62 nm (blue line), respectively. (d) Normalized spectral localized energies of $l_0$=2.16 nm and 1.62 nm in triangular region of MoSe₂ nanodots, respectively. (e) Phonon spectral transmission coefficients of $l_0$=2.16 nm and 1.62 nm, respectively.

To further visualize the transition from incoherent to coherent phonon transport in PAHs, the spatial energy distributions are shown in **Figure 4a-c**. The spatial energy distribution is defined as[18, 46]

$$E_i = \sum_\omega \sum_\lambda \sum_\alpha \left( n + \frac{1}{2} \right) \hbar \omega \varepsilon_{i\alpha,\lambda}^* \, \varepsilon_{i\alpha,\lambda} \delta(\omega - \omega_\lambda) \qquad (3)$$

where $E_i$ is the local energy for the atom $i$ and $n$ is the phonon occupation number given by the Bose-Einstein distribution, which is implemented by the GULP[33]. To present the localized mode, only phonons with participation ratio $P_\lambda < 0.2$ were included in the summation in Eq. (3). The colors from blue to red denotes the increasing localized phonon energy. Obviously, the spatial energy distribution of $MoS_2/MoSe_2$ PAHs is much larger than pristine $MoS_2$, especially in the triangular region of $MoSe_2$ nanodots, which indicates the existence of stronger phonon localization in $MoS_2/MoSe_2$ PAHs. Interestingly, the phonon correlation is rebuilt as the $l_0$ decreases, inducing the coherent phonons dominate the phonon transport at $MoSe_2$ region. In other words, coherent phonon transport is observed in the $MoS_2/MoSe_2$ PAHs of $l_0 \leq 1.62$ nm, which direct visual the crossover from incoherent to coherent phonons. For clarity, the normalized spectral localized energies of the triangular region of $MoSe_2$ nanodots are shown in **Figure 4d**, which exhibit a visible difference between $l_0$=2.16 nm and $l_0$=1.62 nm at middle-low frequency. It is observed that the localized energy of $l_0$=2.16 nm is stronger than that of $l_0$=1.62 nm at 5-8.5 THz, but most of the localized energy of $l_0$=2.16 nm at high frequency is quite weak. This is mainly due to the high-frequency phonons of PAHs of $l_0$=2.16 nm trapped in the $MoS_2$ region (details in supplementary information **Figure S2**), while the low-frequency phonons blocked in the $MoSe_2$ region. In contrast, with the further reducing period length $l_0$, the localized energy of low-frequency phonons becomes weaker, which reveals the underlying reason of rebuilt coherent phonons.

Above the harmonic analysis is not enough to thoroughly understand the result due to the lack of critical information (high order phonon scattering). To reveal the anharmonic effect of phonon transport in PAHs, the spectral phonon transmission coefficients $T(\omega)$ including the whole orders of anharmonicity were calculated by MD simulations, which can be expressed as[18]

$$T(\omega) = \frac{q(\omega)}{k_B \Delta T} \tag{4}$$

where $k_B$ and $\Delta T$ are the Boltzmann parameter and the temperature difference between the two Langevin thermostats, respectively. Here $q(\omega)$ is the frequency-dependent heat current across the imaginary cross-section, which can be calculated as[18]

$$q(\omega) = \frac{2}{t_s} Re \sum_{i \in L} \sum_{j \in R} < \boldsymbol{F}_{ij}(\omega) \cdot \boldsymbol{V}_i(\omega)^* > \tag{5}$$

where $t_s$ is the simulation time, and $\boldsymbol{F}_{ij}$ is the inter-atomic force on atom $i$ due to atom $j$. Here, "L" and "R" denotes the left and right segment, respectively, which are located at two sides of the imaginary cross-section.

The spectral phonon transmission coefficients of MoS$_2$/MoSe$_2$ PAHs with $l_0$=2.16 nm (red line) and 1.62 nm (blue line) are shown in **Figure 4e**, respectively. The calculated results indicate that the phonon transmission coefficients of MoS$_2$/MoSe$_2$ PAHs of $l_0$=2.16 nm and 1.62 nm are significantly lower that of pristine MoS$_2$ at all frequency range (details shown in supplementary information **Figure S3a**). Especially, the phonon transmission coefficients of low-frequency phonon dominating the heat conduction are suppressed, which shows the phonon localization of PAHs is mainly at low-frequency range. In addition, the phonon transmission coefficient of MoS$_2$/MoSe$_2$ PAHs with $l_0$=1.62 nm is larger than that of $l_0$=2.16 nm at around 5-8 THz, indicating the coherent phonons are dominating the phonon transport again. The transmission accumulations (details shown in supplementary information **Figure S3b**) of MoS$_2$/MoSe$_2$ PAHs of $l_0$=2.16 nm and 1.62 nm are obtained by integrating transmission coefficients, which indicate the capacity of thermal transport. Therefore, the differences in transmission accumulation (**Figure S3b**) of among pristine MoS$_2$, MoS$_2$/MoSe$_2$ PAHs of $l_0$=2.16 nm and 1.62 nm are similar with that of the thermal conductivity (**Figure 2a**). However, it is observed that the difference of transmission accumulations between MoS$_2$/MoSe$_2$ PAHs of $l_0$=2.16 nm and 1.62 nm is not much pronounced before 6 THz. This is because that the phonon-interface scattering strengthens caused by more interfaces generated as $l_0$ decreasing, which reduces the phonon mean free path and suppresses the reconstruction of coherent phonon transport.

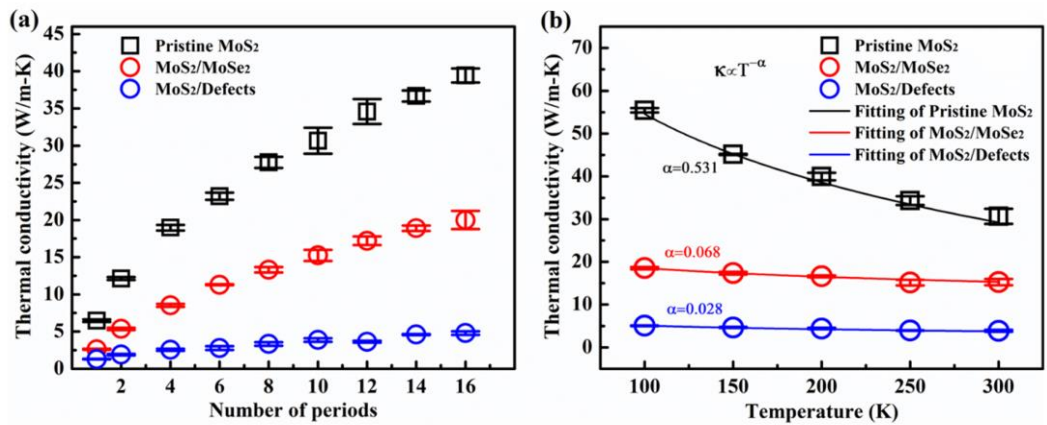

**Figures 5.** (Color online) (a) Thermal conductivities of pristine $MoS_2$, $MoS_2/MoSe_2$ and $MoS_2$/defect PAHs as a function of the number of periods. (b) Thermal conductivities of pristine $MoS_2$, $MoS_2/MoSe_2$ and $MoS_2$/defect PAHs as a function of temperature.

In addition to the physical insights of transition from incoherent to coherent phonon, the length and temperature dependence of $MoS_2/MoSe_2$ PAHs are also investigated in **Figure 5**. The system length is increased by increasing the number of periods when the period length $l_0$ is fixed at 5.40 nm. The thermal conductivities of $MoS_2/MoSe_2$ and $MoS_2$/defect PAHs increase monotonically with the number of periods and gradually converges are shown in **Figure 5a**. Note that, the thermal conductivity of the pristine $MoS_2$ is changed from 6.47 to 39.43 W/m-K, while the thermal conductivity value for $MoS_2/MoSe_2$ ($MoS_2$/defect) PAHs increases from 2.63 (1.30) to 20 (4.80) W/m-K, respectively. It is observed that the thermal conductivity is strongly affected by the system length if the phonon mean-free path is longer than the sample length. Then, the temperature dependences of thermal conductivity of pristine $MoS_2$, $MoS_2/MoSe_2$ and $MoS_2$/defect PAHs from 100 to 300 K are shown in **Figure 5b**. The system length and period length $l_0$ are fixed at 54.02 nm and 5.40 nm, respectively. Thermal conductivities of $MoS_2/MoSe_2$ and $MoS_2$/defect PAHs are insensitive to temperature compare with that of pristine $MoS_2$. Similar temperature dependence is also found in other superlattices[47] and phononic crystals[18]. To further explore the mechanism of phonon-boundary scattering, the 2D temperature distribution[48] was calculated by dividing the $MoS_2/MoSe_2$ PAHs into 3200 equally sized 2D chunks (details in supplementary informational **Figure S8**), which directly visualizes the thermal resistance mapping in PAHs system. It is found that the larger temperature difference at the boundaries of system and $MoSe_2$-nanodots region can be observed due to the phonon-boundary scattering and phonon localization. In general, the normal phonon-phonon scatterings (crystal momentum conserved in three-phonon interactions) and Umklapp phonon-phonon scatterings (crystal momentum not conserved) exist in perfect crystalline. At sufficiently high temperatures, Umklapp scattering is dominated and the thermal conductivity decreases as $\kappa \propto T^{-\alpha}$ ( $\alpha$ =1). However, the power exponent ( $\alpha$ =0.531) of temperature dependence of pristine $MoS_2$ of our MD is less than 1. The mainly originates from the boundary scattering introduced by the size effect when the size of the simulation cell is much smaller than the effective mean free path of pristine $MoS_2$. Similar result is also found in other nanostructured 2D materials[18, 49, 50]. In addition, the exponents of the temperature dependence of $MoS_2/MoSe_2$ and $MoS_2$/defect PAHs are smaller than that of pristine $MoS_2$, which are 0.068 and 0.028, respectively. Therefore, with the increasing of temperature, besides the phonon-phonon scattering, phonon-boundary scattering and phonon-defect scattering dominates in $MoS_2/MoSe_2$ and $MoS_2$/defect PAHs.

**Conclusion**

To summarize, this work investigates the thermal transport properties in single-layer $MoS_2/MoSe_2$ PAHs using NEMD simulations. It is found that a non-monotonic thermal conductivity of $MoS_2/MoSe_2$ PAHs with increasing of the period number, which is due to the phonon transport transition mechanism from incoherent to coherent phonons. The reduction of thermal conductivity with increasing of the period number is due to phonon localizations induces by the vibrational mismatch between embedded nanodots $MoSe_2$ and $MoS_2$ region, which is visualized by spatial energy distribution, mode weight factors and eigenvectors. Furthermore, with further increasing of the period number, the uptrend of thermal conductivity originates from the emergency of coherence phonons in $MoS_2/MoSe_2$ PAHs, which is visually analyzed by phonon dispersion relations, group velocities, and directly visualized based on spatial energy distributions and spectral phonon transmission coefficients. In addition to the study of incoherent and coherent mechanism, the size effects and temperature dependence of $MoS_2/MoSe_2$ PAHs also are presented. This study provides valuable insights into understanding and optimizing thermal management and the thermoelectric performance of superlattice nanodevices.

**Acknowledgement**

This work was supported in part by National Natural Science Foundation of China (Grant No. 52006130 and 11747109), China Postdoctoral Science Foundation (2020M670321), Natural Science Foundation of Shaanxi Province (2020JQ-692), Natural Science Research Start-up Fund of Shaanxi University of Science and Technology (2018GBJ-10), Tsinghua-Foshan Innovation Special Fund (2019THFS0125). The work was carried out at LvLiang Cloud Computing Center of China, and the calculation were performed on Tianhe-2.